\documentclass[oldversion]{aa}
\usepackage{epsfig}
\usepackage{natbib}
\usepackage{lscape}
\usepackage{wasysym}
\bibpunct{(}{)}{;}{a}{}{,}
\begin{document}

\title{Metallicities for 13 nearby open clusters from high-resolution spectroscopy of dwarf and giant stars\thanks{Based on 
observations collected at the La Silla Parana Observatory,
ESO (Chile) with the UVES spectrograph at the 8.2-m Kueyen telescope, under
programs 079.C-0131 and 66.D-0457.}}

\subtitle{Stellar metallicity, stellar mass, and giant planets}

\author{
  N.C. Santos\inst{1} \and	
  C. Lovis\inst{2} \and	
  G. Pace\inst{1} \and
  J. Melendez\inst{1} \and
  D. Naef\inst{3}
  }

\institute{
    Centro de Astrof{\'\i}sica, Universidade do Porto, Rua das Estrelas, 
    4150-762 Porto, Portugal
    \and
    Observatoire de Gen\`eve, 51 ch. des Maillettes, 1290 Sauverny, Switzerland
    \and
    European Southern Observatory, Casilla 19001, Santiago 19, Chile
}


\date{Received 6 October 2008; accepted 14 November 2008}

\abstract{We present a study of accurate stellar parameters and iron abundances
for 39 giants and 16 dwarfs in the 13 open clusters
\object{IC2714}, 
\object{IC4651},  
\object{IC4756},  
\object{NGC2360}, 
\object{NGC2423},
\object{NGC2447} (M93), 
\object{NGC2539}, 
\object{NGC2682} (M67), 
\object{NGC3114}, 
\object{NGC3680}, 
\object{NGC4349}, 
\object{NGC5822}, 
\object{NGC6633}.
The analysis was done using a set of high-resolution and high-S/N spectra
obtained with the UVES spectrograph (VLT). These clusters are currently
being searched for planets using precise radial velocities. For all the
clusters, the derived average metallicities are close to solar. Interestingly,
the values derived seem to depend on the line-list used. This 
dependence and its implications for the study of chemical abundances in giants stars are discussed. 
We show that a careful choice of the lines may be crucial for the derivation 
of metallicities for giant stars on the same metallicity scale as those derived for dwarfs.
Finally, we discuss the implications of the derived abundances for the
metallicity- and mass-giant planet correlation. We conclude that a good knowledge
of the two parameters is necessary to correctly disentangle
their influence on the formation of giant planets.
  }

\authorrunning{Santos et al.}
\titlerunning{Chemical abundances in 13 open clusters}
\maketitle

\section{Introduction}

Precise spectroscopic studies 
of solar-type dwarfs with giant planets have shown that the frequency of giant planets is a 
strong function of the stellar metallicity. It seems easier to find 
a giant planet around a metal-rich star than around a metal-poor object 
\citep[][]{Gonzalez-2001,Santos-2001,Santos-2004b,Santos-2005a,Reid-2002,Fischer-2005}\footnote{For 
a recent general review of the properties of the known exoplanets see \citet[][]{Udry-2007}}. 
This observation has often been interpreted as
being due to a higher probability of forming a giant planet in
a metal-rich disk \citep[e.g.][]{Ida-2004b,Mordasini-2007}.


Interestingly, it has been pointed out that this metallicity-giant planet correlation
may not hold for intermediate mass stars hosting giant planets \citep[][]{Pasquini-2007}.
Pasquini et al. interpret this observation as due to the deeper convective envelopes of
giant stars, which could have washed out the previously existing metallicity enhancement 
due to the accretion of planetary material.
Although not fully accepted \citep[][]{Hekker-2007}, if confirmed this lack
of correlation could alternatively be hinting that
stellar mass strongly influences the planet formation process. 

A few studies found evidence that the fraction of stars 
harboring planets is indeed an increasing function of stellar 
mass \citep[][]{Lovis-2007,Johnson-2007}. For instance, the frequency
of giant planets orbiting (low-mass) M-dwarfs is considerably lower than the
one found for FGK dwarfs \citep[][]{Bonfils-2005b,Endl-2006}. 
This result is expected from the models of planetary formation 
\citep[][]{Laughlin-2004,Ida-2005,Kennedy-2008}, although a consensus does not exist regarding
this point \citep[][]{Kornet-2005,Boss-2006}.

\begin{table}[t]
\caption{Target list of UVES-VLT program 079.C-0131.}
\label{tab:target}
\begin{tabular}{lllr}
\hline\hline
\noalign{\smallskip}
Star  & $\alpha$ (2000.0) & $\delta$ (2000.0) & V \\
\hline
\multicolumn{4}{l}{Giant stars}\\
NGC2360No7    & 07:18:19.01 & -15:34:59.4 & 11.10    \\  
NGC2360No79   & 07:17:16.91 & -15:38:40.7 & 11.29    \\ 
NGC2360No85   & 07:17:32.56 & -15:38:59.4 & 11.44    \\ 
NGC2423No3    & 07:37:09.24 & -13:54:24.0 & 10.08    \\  
NGC2423No20   & 07:37:11.53 & -13:55:44.1 & 11.08    \\  
NGC2423No240  & 07:37:46.60 & -13:49:44.3 & 10.67    \\  
NGC2447No28   & 07:44:50.25 & -23:52:27.1 & 9.85     \\  
NGC2447No34   & 07:44:33.66 & -23:51:42.2 & 10.12    \\  
NGC2447No41   & 07:44:25.73 & -23:49:53.0 & 10.03    \\ 
NGC2539No229  & 08:10:33.80 & -12:51:48.9 & 11.17    \\ 
NGC2539No346  & 08:10:23.02 & -12:50:43.3 & 10.92    \\ 
NGC2539No502  & 08:11:27.67 & -12:41:06.8 & 11.03    \\ 
NGC2682No164  & 08:51:28.99 & +11:50:33.1 & 10.63    \\ 
NGC2682No266  & 08:51:59.52 & +11:55:04.8 & 10.55    \\ 
NGC2682No286  & 08:52:18.57 & +11:44:26.4 & 10.41    \\
NGC3114No13   & 10:01:17.87 & -60:27:02.6 & 9.09     \\
NGC3114No181  & 10:03:51.49 & -60:03:10.2 & 8.30     \\
NGC3114No273  & 10:02:41.26 & -60:29:03.3 & 9.78     \\
IC2714No53    & 11:18:15.89 & -62:42:04.0 & 11.52    \\
IC2714No87    & 11:17:46.85 & -62:36:49.5 & 11.39    \\
IC2714No110   & 11:17:30.57 & -62:42:27.5 & 11.60    \\
NGC3680No13   & 11:25:16.19 & -43:14:24.0 & 10.82    \\
NGC3680No26   & 11:25:38.07 & -43:16:06.5 & 10.94    \\
NGC3680No41   & 11:25:48.54 & -43:09:52.5 & 10.92    \\
NGC4349No5    & 12:23:45.52 & -61:52:15.5 & 11.51    \\
NGC4349No127  & 12:24:35.47 & -61:49:11.7 & 10.83    \\
NGC4349No168  & 12:24:41.37 & -61:52:03.5 & 11.48    \\
NGC5822No102  & 15:03:49.42 & -54:20:10.7 & 10.82    \\
NGC5822No224  & 15:03:22.85 & -54:26:32.9 & 10.81    \\
NGC5822No438  & 15:05:31.14 & -54:26:20.0 & 10.94    \\
IC4651No7646  & 17:24:41.53 & -49:59:06.4 & 10.31    \\
IC4651No9122  & 17:24:50.09 & -49:56:56.1 & 10.86    \\
IC4651No10393 & 17:24:57.75 & -50:01:32.9 & 10.64    \\
NGC6633No100  & 18:27:54.73 & +06:36:00.3 & 8.30     \\
NGC6633No119  & 18:28:17.64 & +06:46:00.1 & 8.95     \\
NGC6633No126  & 18:28:22.98 & +06:42:29.3 & 8.77     \\
IC4756No38    & 18:37:05.21 & +05:17:31.6 & 9.76     \\
IC4756No42    & 18:37:20.77 & +05:53:43.1 & 9.46     \\
IC4756No125   & 18:39:17.88 & +05:13:48.8 & 9.29     \\
\hline
\multicolumn{4}{l}{Dwarf stars}\\
NGC2447\,13625  &  07:44:10.23  &  -24:05:46.1  &  15.36 \\
NGC2447\,13815  &  07:44:11.43  &  -23:53:28.3  &  14.88 \\
NGC2447\,18606  &  07:44:41.97  &  -23:59:14.1  &  15.00 \\
\hline
\noalign{\smallskip}
\end{tabular}
\end{table}

One way to address these issues is to search for planets around (intermediate-mass) 
giant stars. The search for planets around such objects is motivated by some
of the limitations of the radial-velocity technique. Although very effective 
to search for planets around solar-type stars,
for main-sequence higher mass objects this method is usually not
very efficient. This is due to the lack of spectral information (lines),
together with the usually high rotation velocities of higher mass dwarfs.

Intermediate mass stars evolving out of the main sequence, however, become
cooler and (usually) slower rotators. This fact makes of them good targets
for radial-velocity searches, and give the possibility to use such
techniques to probe the frequency of planets around higher mass stars 
(1.5-4M$_\odot$).

A few teams started radial-velocity programs to
search for planets around sub-giant or giant field stars. As a consequence, a few giant planet
candidates have been announced around these kind of objects \citep[e.g.][]{Frink-2002,Sato-2003,Setiawan-2005,Hatzes-2006,Niedzielski-2007,Johnson-2007b}.
Unfortunately, it is not easy to derive the
mass for a field giant star, making it difficult to study the ``stellar mass-frequency of planets" 
relation for these objects.

One way around this difficulty is to search for planets orbiting
giants in galactic open clusters. Three giant planets were detected so far by such surveys, namely orbiting
\object{NGC2423No3}, \object{NGC4349No127} \citep[][]{Lovis-2007}, and \object{$\epsilon$\,Tau} \citep[][]{Sato-2007}.
In this case, the derivation of the mass of the turnoff stars in the cluster gives us immediate 
information about the mass of the clump giants. Furthermore, 
if we observe clusters with different ages we will be studying stars 
within a whole range of stellar masses. 

In this paper we derive atmospheric parameters and chemical abundances
for giant and dwarfs stars in 13 clusters that are being monitored 
with radial-velocities to search for giant planets \citep[][]{Lovis-2007}. 
In Sects.\,\ref{sec:sample} and \ref{sec:observations} we present our sample
and the observations. In Sect\,\ref{sec:analysis} we present the
analysis of the data and compare the metallicities derived from giants and dwarf
stars for the same cluster. In Sects\,\ref{sec:metal} and \ref{sec:planets}
we present the results, discussing in particular their implications for
the metallicity- and mass-giant planet correlations. We conclude 
in Sect\,\ref{sec:conclusion}.

\begin{table*}
\caption[]{Stellar parameters derived for the giants stars using the line-list 
of \citet[][]{Sousa-2008} (S08). See text for more details.}
\begin{tabular}{lcccccc}
\hline
Star     & T$_{\mathrm{eff}}$ & $\log{g}_{spec}$ & $\xi_{\mathrm{t}}$ & [Fe/H] & N(\ion{Fe}{i},\ion{Fe}{ii}) & $\sigma$(\ion{Fe}{i},\ion{Fe}{ii})  \\
   & [K]                &  [cm\,s$^{-2}$]  &  [km\,s$^{-1}$]    &        &                             &                                     \\
\hline
IC2714No53    & 5211$\pm$ 36 & 2.89$\pm$0.36 & 1.64$\pm$0.04 & 0.02$\pm$0.11  & 181,25 & 0.11,0.16  \\
IC2714No87    & 5253$\pm$ 38 & 3.03$\pm$0.31 & 1.69$\pm$0.04 & 0.03$\pm$0.12  & 179,24 & 0.11,0.14  \\
IC2714No110   & 5128$\pm$ 31 & 2.92$\pm$0.27 & 1.56$\pm$0.03 & 0.02$\pm$0.10  & 183,24 & 0.10,0.13  \\
IC4651No7646  & 4974$\pm$ 34 & 2.80$\pm$0.28 & 1.54$\pm$0.03 & 0.09$\pm$0.10  & 182,23 & 0.10,0.14  \\
IC4651No9122  & 4787$\pm$ 43 & 2.82$\pm$0.35 & 1.52$\pm$0.05 & 0.09$\pm$0.12  & 177,24 & 0.12,0.19  \\
IC4651No10393 & 4903$\pm$ 43 & 2.79$\pm$0.30 & 1.33$\pm$0.04 & 0.10$\pm$0.12  & 180,25 & 0.12,0.15  \\
IC4756No38    & 5225$\pm$ 26 & 3.16$\pm$0.22 & 1.40$\pm$0.02 & 0.05$\pm$0.08  & 179,24 & 0.08,0.10  \\
IC4756No42    & 5240$\pm$ 26 & 3.14$\pm$0.22 & 1.39$\pm$0.02 & 0.01$\pm$0.08  & 181,24 & 0.08,0.10  \\
IC4756No125   & 5207$\pm$ 26 & 3.06$\pm$0.26 & 1.47$\pm$0.02 & 0.02$\pm$0.08  & 181,24 & 0.08,0.12  \\
NGC2360No7    & 5182$\pm$ 22 & 3.09$\pm$0.25 & 1.45$\pm$0.02 & -0.00$\pm$0.07 & 181,25 & 0.07,0.12  \\
NGC2360No79   & 5201$\pm$ 25 & 3.07$\pm$0.30 & 1.46$\pm$0.02 & 0.01$\pm$0.08  & 180,25 & 0.08,0.14  \\
NGC2360No85   & 5195$\pm$ 26 & 3.21$\pm$0.29 & 1.40$\pm$0.02 & -0.00$\pm$0.08 & 183,25 & 0.08,0.14  \\
NGC2423No3    & 4703$\pm$ 49 & 2.48$\pm$0.40 & 1.67$\pm$0.05 & -0.00$\pm$0.15 & 180,24 & 0.15,0.21  \\
NGC2423No20   & 5122$\pm$ 33 & 3.09$\pm$0.20 & 1.43$\pm$0.03 & 0.12$\pm$0.09  & 180,24 & 0.09,0.10  \\
NGC2423No240  & 5104$\pm$ 35 & 2.91$\pm$0.20 & 1.51$\pm$0.03 & 0.09$\pm$0.10  & 183,25 & 0.10,0.10  \\
NGC2447No28   & 5125$\pm$ 23 & 2.77$\pm$0.22 & 1.63$\pm$0.02 & -0.06$\pm$0.07 & 174,23 & 0.07,0.10  \\
NGC2447No34   & 5222$\pm$ 24 & 2.95$\pm$0.16 & 1.62$\pm$0.02 & -0.01$\pm$0.08 & 174,24 & 0.08,0.08  \\
NGC2447No41   & 5190$\pm$ 26 & 2.91$\pm$0.23 & 1.62$\pm$0.02 & -0.03$\pm$0.08 & 177,25 & 0.08,0.11  \\
NGC2539No229  & 5145$\pm$ 34 & 3.03$\pm$0.20 & 1.49$\pm$0.03 & 0.09$\pm$0.10  & 183,24 & 0.10,0.10  \\
NGC2539No346  & 5149$\pm$ 33 & 2.97$\pm$0.22 & 1.56$\pm$0.03 & 0.05$\pm$0.10  & 182,25 & 0.10,0.10  \\
NGC2539No502  & 5211$\pm$ 34 & 2.86$\pm$0.47 & 1.60$\pm$0.03 & 0.10$\pm$0.10  & 180,25 & 0.10,0.21  \\
NGC2682No164  & 4812$\pm$ 42 & 2.73$\pm$0.21 & 1.57$\pm$0.04 & 0.03$\pm$0.12  & 180,22 & 0.12,0.11  \\
NGC2682No266  & 4862$\pm$ 37 & 2.76$\pm$0.23 & 1.59$\pm$0.03 & 0.01$\pm$0.11  & 179,23 & 0.11,0.12  \\
NGC2682No286  & 4817$\pm$ 38 & 2.69$\pm$0.23 & 1.59$\pm$0.04 & 0.01$\pm$0.11  & 176,23 & 0.11,0.12  \\
NGC3114No13   & 4985$\pm$ 34 & 2.77$\pm$0.18 & 1.57$\pm$0.03 & 0.01$\pm$0.10  & 181,24 & 0.10,0.09  \\
NGC3114No181  & 4561$\pm$ 45 & 1.92$\pm$0.32 & 2.02$\pm$0.04 & -0.11$\pm$0.13 & 165,24 & 0.13,0.17  \\
NGC3114No273  & 5305$\pm$ 24 & 3.09$\pm$0.20 & 1.57$\pm$0.02 & 0.07$\pm$0.07  & 183,25 & 0.07,0.09  \\
NGC3680No13   & 4781$\pm$ 41 & 2.79$\pm$0.19 & 1.48$\pm$0.04 & -0.01$\pm$0.12 & 181,21 & 0.12,0.10  \\
NGC3680No26   & 4758$\pm$ 50 & 2.70$\pm$0.23 & 1.40$\pm$0.05 & -0.01$\pm$0.14 & 177,23 & 0.14,0.12  \\
NGC3680No41   & 4823$\pm$ 39 & 2.85$\pm$0.25 & 1.45$\pm$0.04 & 0.01$\pm$0.12  & 180,23 & 0.12,0.13  \\
NGC4349No5    & 5186$\pm$ 42 & 2.72$\pm$0.45 & 1.90$\pm$0.04 & -0.00$\pm$0.14 & 180,25 & 0.14,0.21  \\
NGC4349No127  & 4569$\pm$ 69 & 2.08$\pm$0.35 & 1.93$\pm$0.06 & -0.13$\pm$0.18 & 172,23 & 0.17,0.19  \\
NGC4349No168  & 5154$\pm$ 34 & 2.56$\pm$0.34 & 1.88$\pm$0.03 & -0.03$\pm$0.12 & 180,25 & 0.12,0.16  \\
NGC5822No102  & 5253$\pm$ 28 & 3.17$\pm$0.39 & 1.44$\pm$0.03 & -0.00$\pm$0.08 & 178,24 & 0.08,0.17  \\
NGC5822No224  & 5214$\pm$ 28 & 3.14$\pm$0.20 & 1.41$\pm$0.03 & 0.06$\pm$0.08  & 184,25 & 0.08,0.09  \\
NGC5822No438  & 5208$\pm$ 25 & 3.16$\pm$0.19 & 1.39$\pm$0.02 & 0.06$\pm$0.08  & 180,25 & 0.08,0.09  \\
NGC6633No100  & 5118$\pm$ 29 & 2.83$\pm$0.25 & 1.65$\pm$0.03 & 0.04$\pm$0.10  & 183,26 & 0.10,0.12  \\
NGC6633No119  & 5275$\pm$ 30 & 3.10$\pm$0.24 & 1.43$\pm$0.03 & 0.04$\pm$0.09  & 184,24 & 0.09,0.11  \\
NGC6633No126  & 5251$\pm$ 29 & 3.10$\pm$0.29 & 1.50$\pm$0.03 & 0.06$\pm$0.09  & 183,24 & 0.09,0.13  \\
\hline
\end{tabular}
\label{tab:giants}
\end{table*}

\section{Sample}
\label{sec:sample}

For statistical and redundancy reasons, we chose to observe 3 clump giants in 
each of the 13 clusters studied in this paper. The clusters were
taken from the sample of \citet[][]{Lovis-2007}: 
\object{IC2714}, 
\object{IC4651},  
\object{IC4756},  
\object{NGC2360}, 
\object{NGC2423},
\object{NGC2447} (M93), 
\object{NGC2539}, 
\object{NGC2682} (M67), 
\object{NGC3114}, 
\object{NGC3680}, 
\object{NGC4349}, 
\object{NGC5822}, 
\object{NGC6633}.
We point the reader to this paper for more details on the sample clusters.

The choice of the specific targets to observe was based on CORALIE and HARPS 
radial-velocity data collected by these authors in their planet-search 
program. The final list of targets (Table\,\ref{tab:target}) includes either stars hosting
a candidate giant planet or stars that are stable in radial-velocity within the measurement 
errors. This allows us to exclude spectroscopic binary stars from our sample.

The majority of the studies about the metallicity-giant planet 
connection are based on the analysis of field dwarfs 
\citep[e.g.][]{Santos-2004b,Santos-2005a,Fischer-2005,Sousa-2008}.
It is thus important to check that the metallicities derived for the 
giants match those derived for the dwarfs of the same cluster.
With this goal, a number of G-dwarfs in 4 of the clusters mentioned
above were also observed. 

For \object{NGC2447}, three dwarfs (Table\,\ref{tab:target}) were 
chosen based on the 
radial-velocity measurements obtained in a separate planet-search 
project (ESO programs 076.C-0525 and 078.C-0481). This allowed us 
to exclude known spectroscopic binaries, non-members, as well as 
stars with high projected rotational velocities. Details on the
observations and catalog where these stars were selected will
be described in detail in Naef et al. (in prep.).

\begin{table*}
\caption[]{Stellar parameters derived for the giants stars using the line-list 
of \citet[][]{Hekker-2007} (HM07). See text for more details.}
\begin{tabular}{lcccccc}
\hline
Star     & T$_{\mathrm{eff}}$ & $\log{g}_{spec}$ & $\xi_{\mathrm{t}}$ & [Fe/H] & N(\ion{Fe}{i},\ion{Fe}{ii}) & $\sigma$(\ion{Fe}{i},\ion{Fe}{ii})  \\
   & [K]                &  [cm\,s$^{-2}$]  &  [km\,s$^{-1}$]    &        &                             &                                     \\
\hline
IC2714No53    & 5045$\pm$ 65 & 2.75$\pm$0.16 & 1.47$\pm$0.11 & -0.05$\pm$0.10 & 16, 5  & 0.08,0.07  \\
IC2714No87    & 5029$\pm$ 70 & 2.62$\pm$0.27 & 1.44$\pm$0.11 & -0.06$\pm$0.11 & 16, 6  & 0.08,0.13  \\
IC2714No110   & 5017$\pm$ 66 & 2.85$\pm$0.18 & 1.43$\pm$0.10 & 0.01$\pm$0.10  & 16, 6  & 0.08,0.08  \\
IC4651No7646  & 4847$\pm$ 84 & 2.61$\pm$0.18 & 1.23$\dagger$ & 0.20$\pm$0.10  & 16, 6  & 0.09,0.06  \\
IC4651No9122  & 4676$\pm$ 83 & 2.88$\pm$0.19 & 1.32$\dagger$ & 0.23$\pm$0.09  & 16, 6  & 0.09,0.05  \\
IC4651No10393 & 4784$\pm$ 76 & 2.70$\pm$0.24 & 1.26$\dagger$ & 0.12$\pm$0.10  & 16, 6  & 0.09,0.11  \\
IC4756No38    & 5151$\pm$ 73 & 3.16$\pm$0.17 & 1.22$\pm$0.11 & 0.08$\pm$0.11  & 16, 6  & 0.08,0.07  \\
IC4756No42    & 5217$\pm$ 89 & 3.21$\pm$0.17 & 1.15$\pm$0.13 & 0.10$\pm$0.11  & 15, 6  & 0.08,0.06  \\
IC4756No125   & 5146$\pm$ 82 & 3.11$\pm$0.18 & 1.31$\pm$0.13 & 0.07$\pm$0.11  & 15, 6  & 0.08,0.07  \\
NGC2360No7    & 5074$\pm$ 67 & 3.05$\pm$0.15 & 1.48$\pm$0.12 & -0.04$\pm$0.10 & 16, 6  & 0.08,0.06  \\
NGC2360No79   & 5122$\pm$ 69 & 3.16$\pm$0.15 & 1.43$\pm$0.11 & 0.03$\pm$0.10  & 15, 6  & 0.07,0.06  \\
NGC2360No85   & 5091$\pm$ 69 & 3.17$\pm$0.14 & 1.37$\pm$0.11 & -0.01$\pm$0.10 & 16, 6  & 0.07,0.05  \\
NGC2423No3    & 4578$\pm$ 96 & 2.49$\pm$0.22 & 1.60$\pm$0.10 & 0.00$\pm$0.11  & 16, 6  & 0.09,0.06  \\
NGC2423No20   & 5034$\pm$ 75 & 3.13$\pm$0.14 & 1.39$\pm$0.10 & 0.12$\pm$0.09  & 15, 6  & 0.07,0.04  \\
NGC2423No240  & 5030$\pm$ 75 & 2.96$\pm$0.13 & 1.50$\pm$0.11 & 0.08$\pm$0.11  & 16, 6  & 0.08,0.04  \\
NGC2447No28   & 5038$\pm$100 & 2.76$\pm$0.15 & 1.74$\pm$0.16 & -0.12$\pm$0.12 & 12, 6  & 0.08,0.04  \\
NGC2447No34   & 5076$\pm$ 75 & 2.88$\pm$0.10 & 1.65$\pm$0.13 & -0.07$\pm$0.11 & 13, 6  & 0.08,0.02  \\
NGC2447No41   & 5064$\pm$ 84 & 2.93$\pm$0.12 & 1.70$\pm$0.14 & -0.10$\pm$0.12 & 13, 6  & 0.08,0.03  \\
NGC2539No229  & 5067$\pm$ 71 & 3.08$\pm$0.13 & 1.47$\pm$0.11 & 0.11$\pm$0.10  & 16, 6  & 0.08,0.04  \\
NGC2539No346  & 5066$\pm$ 72 & 2.93$\pm$0.13 & 1.38$\pm$0.10 & 0.08$\pm$0.11  & 16, 6  & 0.08,0.04  \\
NGC2539No502  & 5061$\pm$ 81 & 2.93$\pm$0.13 & 1.45$\pm$0.11 & 0.07$\pm$0.11  & 15, 5  & 0.08,0.04  \\
NGC2682No164  & 4659$\pm$ 93 & 2.53$\pm$0.20 & 1.51$\pm$0.10 & -0.02$\pm$0.11 & 16, 6  & 0.09,0.05  \\
NGC2682No266  & 4747$\pm$ 68 & 2.62$\pm$0.15 & 1.48$\pm$0.09 & -0.00$\pm$0.10 & 16, 6  & 0.08,0.05  \\
NGC2682No286  & 4721$\pm$ 75 & 2.68$\pm$0.20 & 1.55$\pm$0.09 & 0.02$\pm$0.09  & 16, 6  & 0.07,0.08  \\
NGC3114No13   & 4871$\pm$ 60 & 2.60$\pm$0.14 & 1.45$\pm$0.08 & 0.01$\pm$0.08  & 15, 6  & 0.06,0.06  \\
NGC3114No181  & 4384$\pm$147 & 1.65$\pm$0.36 & 2.01$\pm$0.15 & -0.23$\pm$0.17 & 16, 6  & 0.14,0.08  \\
NGC3114No273  & 5193$\pm$ 66 & 3.00$\pm$0.16 & 1.50$\pm$0.11 & 0.07$\pm$0.10  & 15, 6  & 0.07,0.07  \\
NGC3680No13   & 4657$\pm$ 74 & 2.68$\pm$0.17 & 1.38$\pm$0.07 & -0.01$\pm$0.08 & 16, 6  & 0.06,0.05  \\
NGC3680No26   & 4649$\pm$125 & 2.68$\pm$0.28 & 1.41$\pm$0.14 & -0.02$\pm$0.14 & 15, 6  & 0.11,0.08  \\
NGC3680No41   & 4676$\pm$ 75 & 2.70$\pm$0.17 & 1.38$\pm$0.08 & -0.03$\pm$0.09 & 16, 6  & 0.07,0.05  \\
NGC4349No5    & 4970$\pm$ 78 & 2.54$\pm$0.17 & 1.69$\pm$0.13 & -0.09$\pm$0.12 & 16, 5  & 0.09,0.07  \\
NGC4349No127  & 4394$\pm$105 & 1.91$\pm$0.31 & 1.81$\pm$0.10 & -0.14$\pm$0.12 & 16, 6  & 0.10,0.13  \\
NGC4349No168  & 5092$\pm$ 54 & 2.76$\pm$0.16 & 1.75$\pm$0.09 & -0.00$\pm$0.08 & 15, 5  & 0.06,0.07  \\
NGC5822No102  & 5170$\pm$ 42 & 3.20$\pm$0.13 & 1.18$\pm$0.07 & 0.05$\pm$0.06  & 13, 6  & 0.04,0.06  \\
NGC5822No224  & 5237$\pm$ 65 & 3.37$\pm$0.12 & 1.15$\pm$0.08 & 0.22$\pm$0.08  & 14, 6  & 0.06,0.04  \\
NGC5822No438  & 5148$\pm$ 62 & 3.21$\pm$0.11 & 1.13$\pm$0.08 & 0.18$\pm$0.08  & 16, 6  & 0.06,0.04  \\
NGC6633No100  & 4979$\pm$ 72 & 2.75$\pm$0.12 & 1.58$\pm$0.10 & 0.00$\pm$0.10  & 16, 6  & 0.08,0.03  \\
NGC6633No119  & 5163$\pm$ 81 & 2.97$\pm$0.12 & 1.25$\pm$0.13 & 0.01$\pm$0.12  & 16, 6  & 0.09,0.04  \\
NGC6633No126  & 5124$\pm$ 69 & 2.92$\pm$0.18 & 1.45$\pm$0.12 & 0.00$\pm$0.11  & 16, 6  & 0.08,0.08  \\
\hline
\end{tabular}
\newline
$\dagger$ Fixed using a T$_{\mathrm{eff}}$-$\xi_{\mathrm{t}}$ calibration (see text)
\label{tab:giantsHM07}
\end{table*}

For the dwarfs in \object{IC4651}, \object{NGC2682}, and \object{NGC3680}, the spectra
used were presented in \citet[][]{Pace-2008} as part of a study of 
chemical abundances in 4 different open clusters. As we will see
below, the metallicity values derived here are in perfect
agreement with those derived by these authors (using the same spectra
but a different methodology).

\begin{table*}
\caption[]{Stellar parameters derived for the dwarf stars. See text for more details.}
\begin{tabular}{lcccccc}
\hline
Star     & T$_{\mathrm{eff}}$ & $\log{g}_{spec}$ & $\xi_{\mathrm{t}}$ & [Fe/H] & N(\ion{Fe}{i},\ion{Fe}{ii}) & $\sigma$(\ion{Fe}{i},\ion{Fe}{ii})  \\
   & [K]                &  [cm\,s$^{-2}$]  &  [km\,s$^{-1}$]    &        &                             &                                     \\
\hline
IC4651\,AMC1109      & 6075$\pm$ 32 & 4.54$\pm$0.22 & 1.14$\pm$0.04 & 0.15$\pm$0.06  & 168,24 & 0.06,0.08  \\
IC4651\,AMC2207      & 6139$\pm$ 34 & 4.58$\pm$0.16 & 1.17$\pm$0.04 & 0.17$\pm$0.06  & 166,22 & 0.06,0.06  \\
IC4651\,AMC4220      & 5793$\pm$ 45 & 4.40$\pm$0.26 & 0.97$\pm$0.06 & 0.15$\pm$0.10  & 169,23 & 0.09,0.10  \\
IC4651\,AMC4226      & 5862$\pm$ 59 & 4.31$\pm$0.46 & 0.89$\pm$0.07 & 0.13$\pm$0.13  & 172,23 & 0.12,0.17  \\
IC4651\,Eggen45        & 6280$\pm$ 44 & 4.40$\pm$0.16 & 1.30$\pm$0.05 & 0.11$\pm$0.10  & 167,22 & 0.09,0.06  \\
NGC2447\,13625 & 5595$\pm$ 30 & 4.45$\pm$0.19 & 1.16$\pm$0.04 & -0.12$\pm$0.06 & 170,21 & 0.06,0.08  \\
NGC2447\,13815 & 5920$\pm$ 28 & 4.59$\pm$0.10 & 1.25$\pm$0.04 & -0.09$\pm$0.05 & 164,21 & 0.05,0.04  \\
NGC2447\,18606 & 5795$\pm$ 33 & 4.47$\pm$0.20 & 1.16$\pm$0.05 & -0.11$\pm$0.07 & 167,22 & 0.07,0.08  \\
NGC2682\,Sanders746         & 5703$\pm$ 30 & 4.46$\pm$0.15 & 0.97$\pm$0.04 & -0.07$\pm$0.06 & 168,22 & 0.06,0.06  \\
NGC2682\,Sanders1048        & 5915$\pm$ 35 & 4.48$\pm$0.32 & 0.96$\pm$0.05 & 0.07$\pm$0.07  & 170,24 & 0.07,0.12  \\
NGC2682\,Sanders1092        & 6074$\pm$ 29 & 4.39$\pm$0.08 & 1.35$\pm$0.03 & 0.04$\pm$0.06  & 167,21 & 0.06,0.03  \\
NGC2682\,Sanders1255        & 5892$\pm$ 23 & 4.47$\pm$0.21 & 1.23$\pm$0.03 & 0.01$\pm$0.05  & 165,26 & 0.05,0.08  \\
NGC2682\,Sanders1283        & 6028$\pm$ 29 & 4.42$\pm$0.13 & 1.15$\pm$0.03 & 0.00$\pm$0.06  & 168,22 & 0.06,0.05  \\
NGC2682\,Sanders1287        & 6111$\pm$ 42 & 4.47$\pm$0.26 & 1.09$\pm$0.05 & 0.01$\pm$0.09  & 161,21 & 0.08,0.09  \\
NGC3680\,AHTC1009    & 5932$\pm$ 28 & 4.47$\pm$0.16 & 1.06$\pm$0.04 & -0.03$\pm$0.06 & 167,24 & 0.06,0.06  \\
NGC3680\,Eggen70       & 6134$\pm$ 42 & 4.54$\pm$0.22 & 1.22$\pm$0.06 & -0.04$\pm$0.09 & 166,22 & 0.08,0.08  \\
\hline
\end{tabular}
\label{tab:dwarfs}
\end{table*}

Finally, given the very good agreement between the values determined by
\citet[][]{Pace-2008} and the ones presented here for the same stars,
the average metallicity values presented by Pace et al. 2009 (in preparation)
for dwarfs in \object{IC4756} and \object{NGC5822} were also added 
to the list.

\section{Observations}
\label{sec:observations}

Spectra for all the 39 giants in the 13 clusters, as well as for 
the three G-dwarfs in \object{NGC2447}, were obtained using the 
UVES spectrograph (VLT Kueyen telescope), between April and October 2007
(ESO programs 079.C-0131A and B). For consistency, we opted for the
same instrument configuration for all the observations.
A slit width of 0.9 arcsec was used, providing a spectral resolution 
R=$\lambda/\Delta\lambda\sim$50\,000. The observations
were done using the RED580 mode, and the resulting spectra 
cover the wavelength domain between 4780 and 6805\AA, with 
a gap between 5730 and 5835\AA\ (corresponding to the gap in the
CCD mosaic). The S/N of the final spectra
is between 200 and 300 for the giants, and $\sim$100 for the dwarfs.

For some of the stars, attention was given to the orientation of
the UVES slit due to the relative crowdedness of the fields. 

As mentioned above, for \object{IC4651}, \object{NGC2682}, 
and \object{NGC3680}, the spectra used are presented 
in \citet[][]{Pace-2008}. These data were also obtained with the UVES
spectrograph (ESO run 66.D-0457), and have a spectral resolution of
100\,000 (a slit width of 0.4 arcsec was used). All spectra cover the same
wavelength interval as the data described above. The final spectra
have a typical S/N of $\sim$80. We point the reader to these authors 
for more details about this set of data.

{
Finally, the spectroscopic parameters derived by Pace et al. (2009, in preparation)
and presented in this paper for dwarfs in \object{IC4756} and \object{NGC5822},
were also obtained from the analysis of UVES data. In this case, the
spectra have a resolution of $\sim$100\,000, and S/N$\sim$100. These
spectra cover the same wavelength domain as the remaining data.
}

\section{Analysis}
\label{sec:analysis}

Stellar atmospheric parameters and iron abundances were derived 
in LTE using the 2002 version of the code 
MOOG \citep[][]{Sneden-1973}\footnote{http://verdi.as.utexas.edu/moog.html} 
and a grid of Kurucz Atlas plane-parallel model atmospheres \citep[][]{Kurucz-1993}. 
The final parameters were obtained imposing excitation and ionization 
equilibrium to a set of \ion{Fe}{i} and \ion{Fe}{ii} lines, following
the basic prescription described in \citet[][]{Santos-2004b}.

The determination of the uncertainties in the derived parameters also follow the same 
prescription as in \citet[][]{Santos-2004b}. The error bars only reflect
internal uncertainties. We conservatively 
use the rms around the average abundance given by \ion{Fe}{i} lines, and not the error on the average,
to derive the error in [Fe/H]. This means that the tabled
uncertainties are likely upper limits. 

Line equivalent widths (EW) for the iron lines were measured using the
automatic ARES code \citep[][]{Sousa-2007,Sousa-2008}\footnote{This software 
can be downloaded at http://www.astro.up.pt/$\sim$sousasag/ares/}. Star-by-star visual
inspection of sample EW measurements was done in order to verify if the 
ARES parameters used were adequate for the resolution and S/N of the
data.

Slightly different procedures were then adopted for the spectroscopic analysis of the
dwarfs and giants.
In the former case, stellar parameters were derived using the line-list described 
in \citet[][]{Sousa-2008} (hereafter S08). This list, composed of 263 \ion{Fe}{i} and 36 \ion{Fe}{ii} lines
in the optical domain has shown to give excellent results for the analysis 
of dwarf stars. For the giants, further to this we also derived the stellar
parameters using the line-list provided by \citet[][]{Hekker-2007} (hereafter HM07). Although
much smaller (20 \ion{Fe}{i} and 6 \ion{Fe}{ii} lines), the lines
in this list were carefully chosen for the analysis of giant stars, avoiding line-blending
from CN lines \citep[][]{Melendez-1999}.

For the three giants in \object{NGC4651}, the small
EW interval of the \ion{Fe}{i} lines measured
using the smaller line-list of HM07 did
not allow a consistent determination of the microturbulence.
We have thus decided to use, only in this case, microturbulence values
derived using the empirical formula used by these authors (based
on a relation between microturbulence and effective temperature).

The derived stellar parameters and iron abundances for all the stars analyzed
in this paper are listed in Tables\,\ref{tab:giants}, \ref{tab:giantsHM07}, and \ref{tab:dwarfs}.

\subsection{Comparison with other works}
\label{sec:comparison}

The large majority of the clusters studied here do not have any
previous metallicity estimate derived from high-resolution 
spectroscopy. 

\citet[][]{Pace-2008} compiled a long list of
chemical abundances for open clusters studied in
the literature using high resolution spectroscopy. A simple comparison 
of their Tables\,3 and 9 for the three clusters in common (namely \object{IC4651}, 
\object{NGC2682} (M67), and \object{NGC3680}) show a very good agreement,
not only with the values derived by these authors but also with other
literature values.

We note that in this paper we analyzed the same spectra as \citet[][]{Pace-2008} for the dwarfs 
in these three clusters, although we used a different procedure to derive the stellar parameters. Pace et al.
find average [Fe/H] values of $+$0.12$\pm$0.05, $+$0.03$\pm$0.04, and $-$0.04$\pm$0.03 for
\object{IC4651}, \object{NGC2682} (M67), and \object{NGC3680}, respectively.
From the same spectra we derive average metallicities 
of $+$0.15$\pm$0.02, $+$0.01$\pm$0.04, and $-$0.03$\pm$0.01.


\citet[][]{Chen-2003b} compiled metallicity estimates (mostly photometric)
for more than 150 open clusters. Except for \object{NGC2447}, all 
the 13 clusters studied here have metallicity values listed by these authors.
A comparison of their [Fe/H] values with those
derived in our paper reveal a reasonable good agreement. The average difference
between our metallicity values and those in Chen et al. is only $-$0.03\,dex
(median of $-$0.04), our determinations having a higher average value (considering
the estimates based on the S08 line-list). 
The dispersion is, however, relatively high, with the differences ranging 
between $+$0.05 and $-$0.15\,dex.

\subsection{Comparing giants and dwarfs}
\label{sec:giantsdwarfs}

The analysis of both dwarfs and giant stars in 6 of the studied clusters
(IC4651, IC4756, NGC2447, NGC2682, NGC3680, and NGC5822)
allows us to verify if the metallicity scale of the two classes of objects is the
same. 

\begin{table}[t]
\caption[]{Weighted average metallicities of the 3 giant stars in each of the 13 clusters. 
$<$[Fe/H]$>_{S08}$, $<$[Fe/H]$>^c_{S08}$, and $<$[Fe/H]$>_{HM07}$ represent the
average values of the metallicities derived using the \citet[][]{Sousa-2008} line-list
without and with the correction discussed in Sect.\,\ref{sec:giantsdwarfs}, and the
average of the values derived using the line-list of \citet[][]{Hekker-2007}. }
\begin{tabular}{lccc}
\hline
Cluster     & $<$[Fe/H]$>$$_{\rm S08}$ & $<$[Fe/H]$>$$^c_{\rm S08}$ & $<$[Fe/H]$>$$_{\rm HM07}$ \\
\hline
IC2714 &  0.02$\pm$0.01 &  0.01$\pm$0.01 & -0.03$\pm$0.04 \\
IC4651 &  0.09$\pm$0.01 &  0.15$\pm$0.01 &  0.19$\pm$0.06 \\
IC4756 &  0.02$\pm$0.02 &  0.02$\pm$0.02 &  0.08$\pm$0.01 \\
NGC2360 &  0.00$\pm$0.01 & -0.03$\pm$0.01 & -0.01$\pm$0.03 \\
NGC2423 &  0.09$\pm$0.06 &  0.14$\pm$0.06 &  0.07$\pm$0.06 \\
NGC2447 & -0.03$\pm$0.03 & -0.10$\pm$0.03 & -0.10$\pm$0.03 \\
NGC2539 &  0.08$\pm$0.03 &  0.13$\pm$0.03 &  0.09$\pm$0.02 \\
NGC2682 &  0.02$\pm$0.01 &  0.00$\pm$0.01 &  0.00$\pm$0.02 \\
NGC3114 &  0.02$\pm$0.09 &  0.02$\pm$0.09 &  0.00$\pm$0.12 \\
NGC3680 & -0.00$\pm$0.01 & -0.04$\pm$0.01 & -0.02$\pm$0.01 \\
NGC4349 & -0.04$\pm$0.06 & -0.12$\pm$0.06 & -0.06$\pm$0.08 \\
NGC5822 &  0.04$\pm$0.04 &  0.05$\pm$0.04 &  0.12$\pm$0.10 \\
NGC6633 &  0.04$\pm$0.01 &  0.06$\pm$0.01 &  0.00$\pm$0.00 \\
\hline
\end{tabular}
\label{tab:avggiants}
\end{table}

In Tables\,\ref{tab:avggiants} and \ref{tab:avgdwarfs} we list the average
values for the metallicities derived for each cluster, using the giant
and dwarf stars, respectively. For the giants, both determinations
are used, using the S08 and HM07 line-lists (columns 2 and 4 in Table\,\ref{tab:avggiants}). 
In general terms, a good agreement is seen. Values determined
using giant stars and the S08 line-list show a star-to-star dispersion that is clearly smaller
than the one obtained using the HM07 line-list. This result is somewhat 
expected given the much higher number of iron lines used in the former case.

\begin{figure*}[t!]
\resizebox{8.5cm}{!}{\includegraphics{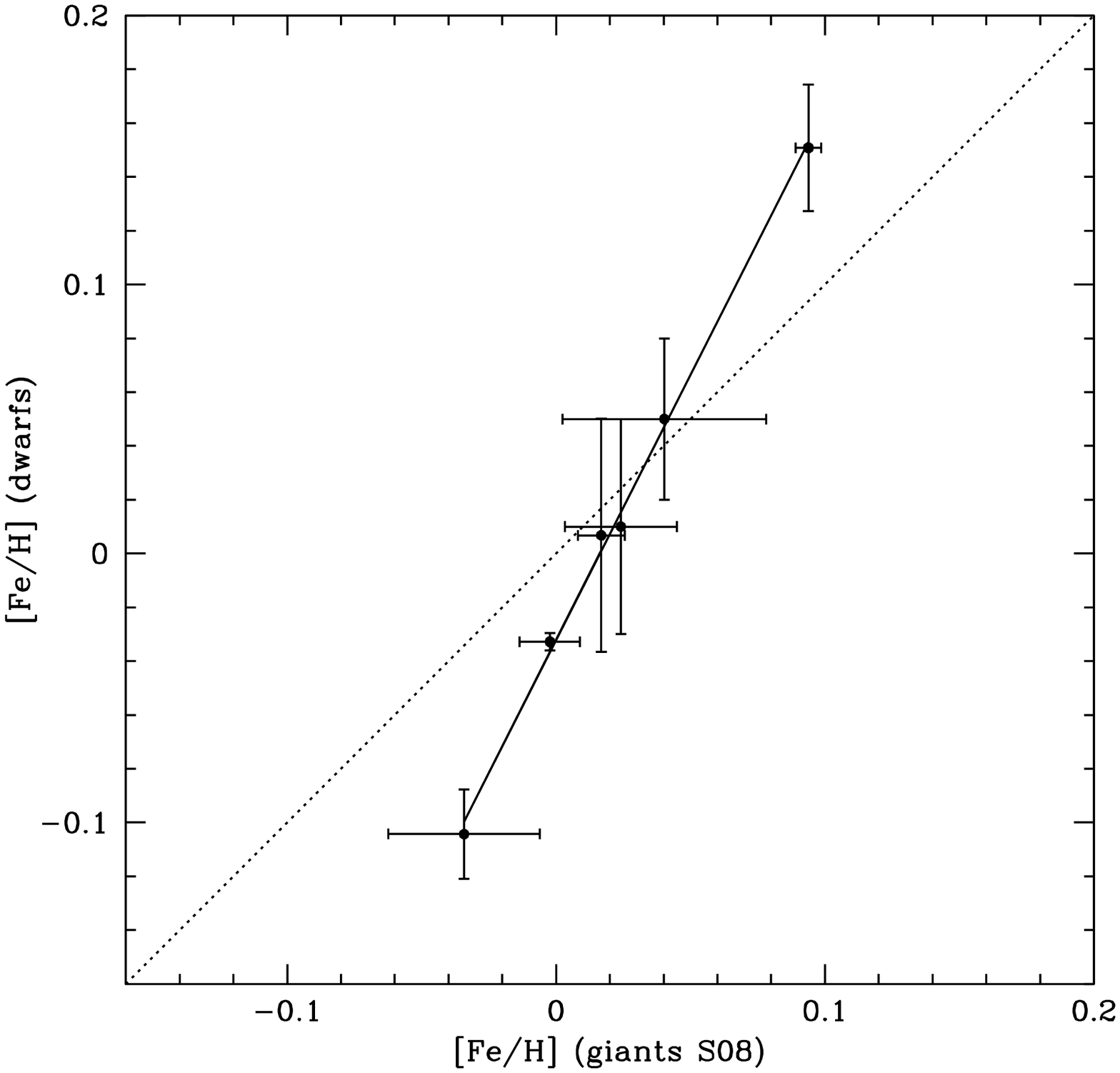}}
\resizebox{8.5cm}{!}{\includegraphics{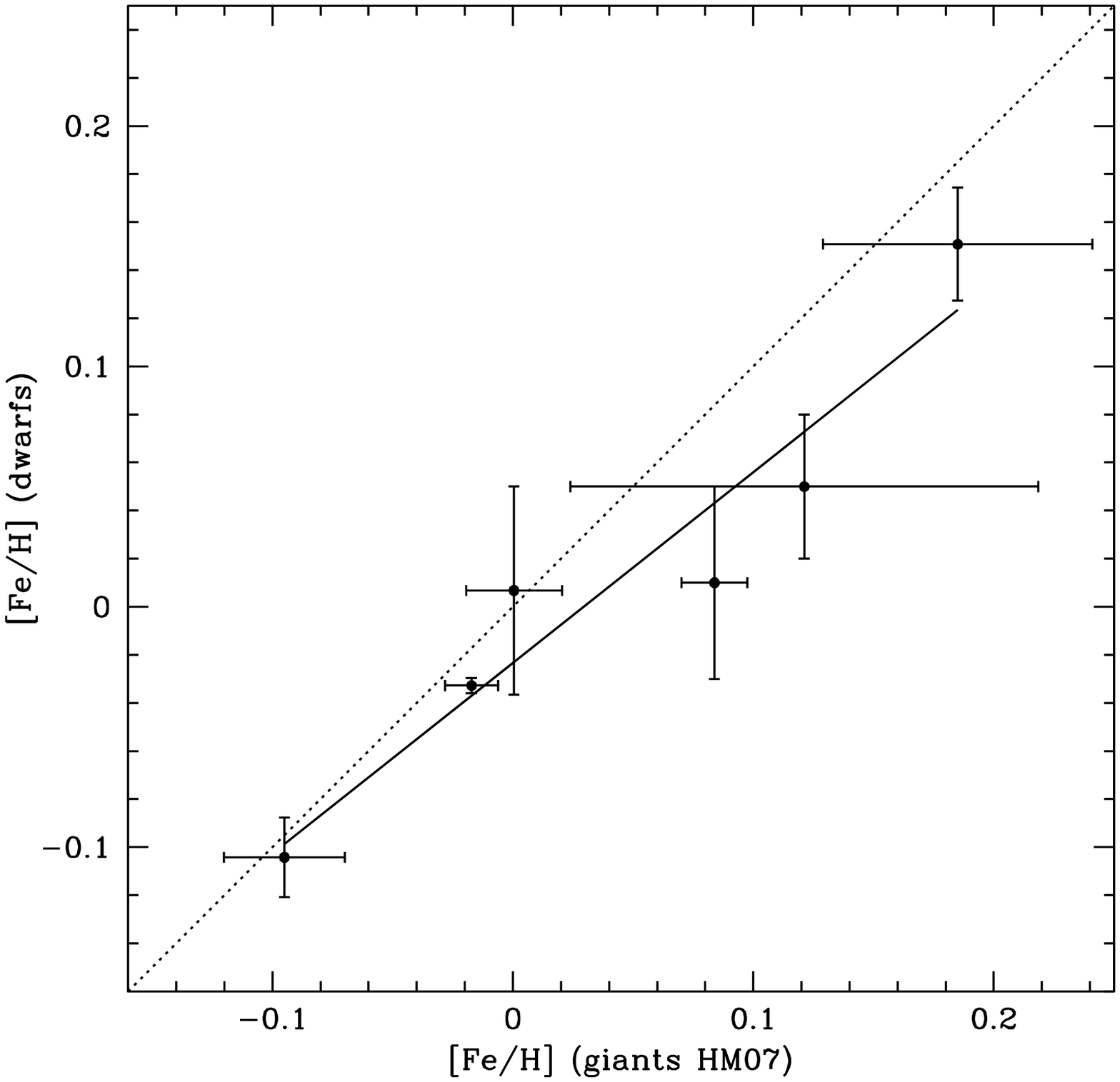}}
\caption{Comparison of the average values for the metallicities derived from
dwarfs and giants in the 6 clusters where analysis was done using both groups of stars.
The dotted line represents the 1:1 relation, while the solid line represents
a linear fit to the points. S08 and HM07 refer to the abundances derived
using the line-lists of \citep[][]{Sousa-2008} and \citep[][]{Hekker-2007},
respectively. The fit in the left panels holds the relation 
$<$[Fe/H]$>_{dwarfs}$=1.97\,$<$[Fe/H]$>_{S08}-$0.03, with an rms of only 0.004\,dex.} 
\label{fig:dwarfsgiants}
\end{figure*}

In Fig\,\ref{fig:dwarfsgiants} we compare the average metallicity values for the
6 clusters derived using the dwarfs stars with the
same values derived using giants, both with the S08 (left panel) and 
HM07 (right panel) line-lists. As mentioned above, in both cases we can see 
a general agreement. However, it is clear from the plots that while
in the latter case there is no evidence of a clear difference on the metallicity
scale for giants and dwarfs, in the former case this does not seem to be true. 
A linear fit to the points in the left panel of Fig\,\ref{fig:dwarfsgiants}
holds a relation $<$[Fe/H]$>_{dwarfs}$=1.97\,$<$[Fe/H]$>_{S08}-$0.03,
with a very small dispersion of 0.004\,dex. 
Interestingly, for near solar metallicities all estimates almost 
perfectly agree. 

This result suggests that the metallicities derived for giant stars using the
HM07 line-list are closer to the expected value, supposing that dwarfs
provide a correct metallicity scale. We remember that the
line-list compiled by HM07 was carefully chosen for the analysis of
giant stars, while the former was optimized for the study of solar-type
dwarfs. 

Given the smaller dispersion of the [Fe/H] values derived for giant stars using
the S08 line-list, it may be interesting, however, to use these values after correcting
for the trend mentioned above. The corrected values are listed
in Table\,\ref{tab:avggiants} (column 3, $<$[Fe/H]$>^c_{S08}$). 

We decided to adopt these values for the rest of the paper.
After correcting the S08 values for the relation mentioned above,
a 1 to 1 relation is also found when comparing with the [Fe/H] 
derived using the HM07 line-list.
We caution, however, that the use of only 6 clusters may limit the 
validity of this correction. More data may be needed to confirm
which of the line-lists provides the best metallicity determinations
for the giant stars.

\begin{table}[t]
\caption[]{Weighted average metallicities of the dwarf stars in each of the
studied clusters.}
\begin{tabular}{lccc}
\hline
Cluster     & $<$[Fe/H]$>$$_{dwarfs}$ & N$_{stars}$ & Source\\
\hline
IC4651 &  0.15$\pm$0.02  & 5 & (1)\\
IC4756 &  0.01$\pm$0.04  & 3 & (2)\\
NGC2447 & -0.10$\pm$0.02 & 3 & (1)\\
NGC2682 &  0.01$\pm$0.04 & 6 & (1)\\
NGC3680 & -0.03$\pm$0.00 & 2 & (1)\\
NGC5822 &  0.05$\pm$0.03 & 2 & (2)\\
\hline
\end{tabular}
\newline
(1) This paper; (2) Pace et al. (in prep.)
\label{tab:avgdwarfs}
\end{table}

We tested the two line-list used, deriving the solar parameters and iron abundance using an
available solar ``Ganymede'' spectrum taken with the HARPS spectrograph\footnote{http://www.eso.org/sci/facilities/lasilla/ instruments/harps/inst/monitoring/sun.html}. In both cases, 
EW were measured with ARES using the parameters set as discussed in S08. As expected, in the former case
the solar parameters almost perfectly match the expected values (T$_\mathrm{eff}$, $\log{g}$, $\xi_t$, [Fe/H] of 
5765$\pm$28\,K, 4.46$\pm$0.11,0.97$\pm$0.04, and $-$0.01$\pm$0.05, respectively). Using the HM07 line-list 
we obtain a slightly different solution (5777$\pm$38\,K, 4.63$\pm$0.12, 1.03$\pm$0.06, $+$0.08$\pm$0.05).

\subsection{The metallicity scale for giants}

The bias mentioned above implies that using the metallicities derived for giants
with the S08 line-list we would obtain a significant narrower [Fe/H] distribution. 
In other words, we would overestimate the [Fe/H] values for metal-poor objects, 
while strongly underestimating the metallicities for any putative high [Fe/H] giants. 

This issue may be related to the fact that (cool) giant stars have higher
macroturbulence velocities \citep[][]{Gray-1992}, and that thousands of molecular lines (CN, C2, CH, MgH) 
contribute to the optical spectra of giants \citep[][]{Coelho-2005}. Both effects lead to stronger line-blending, 
implying a wrong estimate of the continuum position when measuring line-EWs. This effect should
be more important for higher metallicity objects, explaining the stronger difference
observed for the most metal-rich stars. 

{
This effect may be enhanced if the spectral resolution is not sufficiently high. 
Using different resolution degraded solar spectra and the ARES code with fixed input 
parameters (as an approximation), \citet[][]{Sousa-2007} suggest that for spectral 
resolutions below $\sim$50\,000, the EW
estimates may tend to systematically decrease as a function of decreasing resolution. In our case, 
all spectra have a resolution above $\sim$50\,000, likely limiting the importance of this problem.
We cannot exclude, however, that this effect is more important for
cool giant stars, for which line crowdedness is stronger.
}

The problem of the metallicity scale for giants stars was already discussed by 
\citet[][]{Taylor-2005}. These authors further pointed out that no giant stars are 
observed in the solar-neighborhood having a metallicity above $\sim$0.2\,dex. Whether this
observation is real or due to some bias in the spectroscopic analysis
should be addressed. Although more data is needed, the results presented above may support
the second hypothesis. A careful choice of the lines used for the analysis
of giants stars is certainly needed.

Finally, the observed bias may have important implications for the study
of stellar populations, like the galactic bulge, where dwarfs are too faint
to be observed. Indeed, the few studies of the chemical abundances of
dwarfs in the bulge have suggested that these may be more metal rich
than previously estimated from the analysis of giant stars \citep[e.g.][]{Cohen-2008}.

\section{The metallicity of nearby open clusters}
\label{sec:metal}

\begin{figure}[t!]
\resizebox{\hsize}{!}{\includegraphics{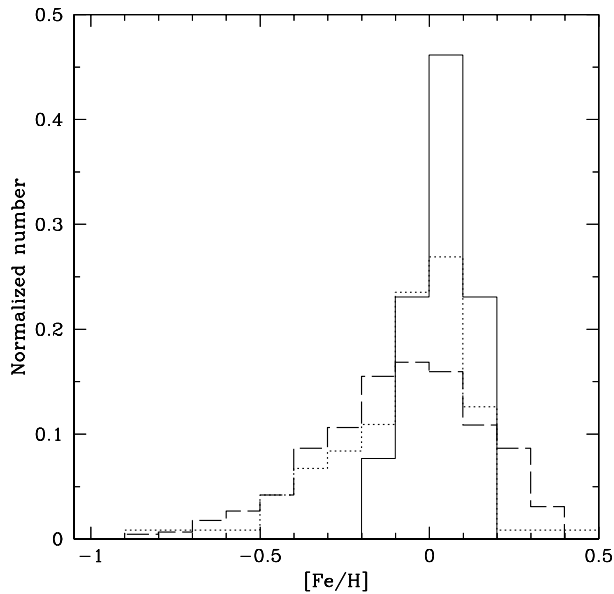}}
\caption{Metallicity distribution of the average metallicities for the
clusters studied in this work using the corrected S08 values listed in Table\,\ref{tab:avggiants}
(filled line), compared with the same distribution for 451 field dwarfs (dashed line) from \citep[][]{Sousa-2008},
and with the [Fe/H] distribution of 119 open clusters (dotted line) described in \citet[][]{Chen-2003}.} 
\label{fig:histo}
\end{figure}

In Fig.\,\ref{fig:histo} we present the metallicity distribution of the 13 clusters
studied in this paper. For this plot we used the average metallicities listed
in Table\,\ref{tab:avggiants}, derived using the S08 line-list and corrected
using the relation discussed in Sect.\,\ref{sec:giantsdwarfs}. A similar
result would be obtained if we were plotting the metallicity values
deriving using the HM07 line-list.

The metallicity distribution of the 13 clusters studied here has a mean value of
$+$0.02\,dex, somehow higher than the value observed for average field dwarfs ($-$0.09\,dex),
and for the open clusters from Table\,1 of \citet[][]{Chen-2003} ($-$0.07\,dex).
The rms of the cluster distribution is also smaller (0.08\,dex, compared to
0.24\,dex for the field dwarfs and 0.23 of the Chen et al. sample).
The average difference with respect to this latter sample is probably not significant, given
the small number of clusters analyzed here, and the possible existence of 
small systematic errors in the metallicities derived using different methods 
(photometry vs. spectroscopy).

\section{Giants with giant planets}
\label{sec:planets}

Two of the giants studied in this work are known to be orbited by
giant planets: \object{NGC2423No3}, \object{NGC4349No127} \citep[][]{Lovis-2007}.
The two clusters have average metallicities of $+$0.14$\pm$0.06 and $-$0.12$\pm$0.06,
respectively (using the corrected values in Table\,\ref{tab:avggiants}).
These two clusters have the second highest, and the lowest metallicity values
among the 13 studied clusters, respectively for \object{NGC2423} and \object{NGC4349}.

The small [Fe/H] interval and the low number of planets discovered so far
in this sample prevents us from taking any clear conclusions regarding the metallicity-giant
planet connection. All the clusters span a [Fe/H] domain where a large number
of planets have been found orbiting dwarf field stars \citep[e.g.][]{Santos-2005a,Fischer-2005,Sousa-2008}.

There is currently some debate about whether the metallicity-giant
planet correlation that is found for field dwarfs \citep[e.g.][]{Santos-2004b,Fischer-2005}
is also valid for giant stars \citep[][]{Pasquini-2007,Hekker-2007}.
Pasquini et al. suggested that this trend in indeed not found for
field giants. In face of our present results, their conclusions should be partially taken with care. 
It could be that their metallicities for metal-rich giants 
have been underestimated, leading to a spurious lack of very metal-rich planet 
hosting giants.

Furthermore, we note that the determination of
accurate stellar masses for field giants is problematic.
In this sense, it is interesting to see that although \object{NGC4349} is among the most metal-poor
clusters analyzed here, the mass of its giants \citep[3.9$\pm$0.3\,M$_\odot$ --][]{Lovis-2007} 
is one of the highest among the sample (see Table 1 of that paper). In comparison,
the giants in the moderately metal-rich cluster NGC2423 are less
massive (2.4$\pm$0.2\,M$_\odot$). 
{If we consider that 
stellar mass positively influences the planet formation efficiency,
this may hint that stellar mass
may be compensating for the lower metallicity in the case of \object{NGC4349},
justifying the discovery of a giant planet among one of its giants.
}

It is curious to see that both \object{NGC2423No3} and \object{NGC4349No127}
are amongst the coolest stars studied in this paper, and the coolest
and more metal-poor objects among the three stars analyzed in their 
respective clusters (although the metallicities all agree within 
the error-bars). We cannot exclude that the lower metallicities 
reflect the existence of systematic errors in the analysis of 
lower-temperature stars, as blending by molecular lines will be more severe 
for these objects \citep[][]{Coelho-2005}. A look at Tables\,\ref{tab:giants} 
and \ref{tab:giantsHM07} show that there is a small tendency 
for lower temperature stars in a given cluster to present also 
lower metallicity values. This possible trend should be studied using
more giant stars covering a wide range of temperatures in each of the 
clusters.

\section{Concluding remarks}
\label{sec:conclusion}

We presented a study of iron abundances in 39 giants and 16 
dwarfs in 13 open clusters using high resolution spectroscopy. 
All these clusters are part of a survey for giant planets 
orbiting intermediate-mass giant stars. 

The results show that all the clusters studied have [Fe/H] values
close to solar, with a small dispersion. Interestingly, the
metallicity values obtained for the giant stars depend on the 
line-list used to derive stellar parameters and iron abundances. 
When comparing with the dwarfs in our sample, we show that
a careful choice of the lines used to derive parameters for giant
stars may be necessary to avoid important biases on the metallicity
scale for these objects. This result may have important implications
for the study of chemical abundances in stellar populations using cool giant stars.

Two of the stars studied (in \object{NGC2423} and \object{NGC4349}) 
are known to be orbited by long period giant planets. NGC2423 is 
one of the most metal-rich clusters in our sample. Interestingly, 
the mass of the giant stars in the most metal-poor of these clusters (NGC4349)
is one of the highest among the 13 clusters. In the context of the
metallicity- and stellar mass-giant planet correlation, this may be hinting that stellar mass
may be compensating for the lower metallicity of this cluster. In any case, this
result shows that to disentangle the metallicity and mass effects
on the frequency of planets we need to have accurate values for
both parameters.

\begin{acknowledgements}
N.C.S. and J.M. would like to thank the support from Funda\c{c}\~ao para a 
Ci\^encia e a Tecnologia, Portugal, through programme Ci\^encia\,2007.
\end{acknowledgements}

\bibliographystyle{aa}
\bibliography{santos_bibliography}

\end{document}